\documentclass[article,epsfig,twocolumn]{article}
\usepackage{graphics}
\usepackage{graphicx}
\usepackage{enumerate} 
\usepackage{amssymb}
\usepackage{xcolor}
\usepackage{amsmath}
\begin{document}
\title{Effect of decreasing population growth-rate on  deforestation and population sustainability}
%\author{ Gerardo Aquino$^{2}$ and Mauro Bologna$^{1}$ }
%\affiliation{$^{1}$Instituto de Alta Investigaci\'{o}n, Universidad de Tarapac\'{a}, Casilla 6-D, Arica, Chile,\\$^{2}$Alan Turing Institute }

\author{Gerardo Aquino and Mauro Bologna}

\date{ Goldsmiths, University of London; aquigerardo@gmail.com\\
Departamento de Ingenier\'ia El\'ectrica-Electr\'onica, Universidad de Tarapac\'a, Arica, Chile; mauroh69@gmail.com}\maketitle

%\affil[+]{these authors contributed equally to this work}7

%\keywords{Keyword1, Keyword2, Keyword3}
%\usepackage{authblk}

\begin{abstract}
We consider the  effect of non-constant parameters  on the   human-forest interaction logistic model coupled with  human technological growth introduced in \cite{noi}.  In recent years in fact, a decrease in  human population growth  rate has emerged which can be measured to about 1.7\%  drop per year since 1960  value which coincides with latest UN projections for next decades up to year 2100  \cite{unp}. We therefore consider here  the effect of decreasing human population growth-rate on the aforementioned model and we evaluate its effect on the probability of survival of human civilisation without going through a catastrophic collapse in population. We find that for  realistic values of  the  human population carrying capacity of the earth (measured by parameter $\beta$)  this decrease would not affect previous  results leading to a low probability of avoiding a catastrophic collapse. For larger more optimistic values of $\beta$  instead, a  decrease in growth-rate would tilt the probability  in favour of a  positive outcome, i.e. from 10-20\%  up to even $95 \%$  likelihood of avoiding collapse. 
\end{abstract}

%\maketitle
% * <john.hammersley@gmail.com> 2015-02-09T12:07:31.197Z:
%
%  Click the title above to edit the author information and abstract
%

%\noindent Please note: Abbreviations should be introduced at the first mention in the main text no abbreviations lists. Suggested structure of main text (not enforced) is provided below.

\section*{Introduction}
The problem of the survival of humanity, for long time the subject of science fiction and catastrophist movies,  has  recently become central in both scientific and social debate, due to various factors, among them climate changes, intensive exploitation of resources and more generally a deterioration of the planetary ecosystem. Recently, the authors of Ref.\cite{noi} pointed out the serious repercussions on the life of the planet of  uncontrolled deforestation. The model examined in \cite{noi} considered the strong connection between the use of the resources (i.e. the forests) and the technological development~\cite{epl,noi}  governed by the following equations

\begin{eqnarray}\label{res-inter2}
\frac{d}{dt}N(t)&=&rN(t)\left[ 1-\frac{N(t)}{\beta R(t)}\right] ,
\\\label{log-inter2}
\frac{d}{dt}R(t)&=&r^{\prime }R(t)\left[ 1-\frac{R(t)}{R_{c}}
\right] -a_0 N(t)R(t) 
\end{eqnarray}
for the  the world population $N$ and the forest-covered surface $R.$ The  parameters involved in Eqs. (1) and (2) are: $\beta$, a positive constant related to the human population carrying capacity of the earth, $r$  the growing rate for humans (estimated as $r\sim 0.01$years$^{-1}$)~\cite{fort}, $a_0$ which may be identified as the technological parameter representing the ability of
exploiting the resources, $r'$ the renewability parameter characterising how quickly the resources are able to regenerate and finally $R_c$, the resource carrying capacity of the earth that in our case may be identified with the initial 60 million square Kilometers of forest.

\begin{equation}
\frac{1}{R}\frac{dR}{dt}\approx -a_0 N
\label{ext}
\end{equation}
The actual population of the earth is $N\sim 7.5 \times 10^{9}$ inhabitants with a maximum carrying capacity estimated~\cite{wil} of $N_c\sim  10^{10}$ inhabitants. The forest carrying capacity may be taken $R_c\sim 6\times 10^7$ Km$^2$ of forest~\cite{rich} while the actual surface of forest is $R\lesssim4\times 10^7$ Km$^2$. We may estimate the $\beta$ parameter as $\beta \sim N_c/R_c\sim 170$ or using actual data of population  growth ~\cite{john}. In this case we obtain a range $ \beta \simeq700$. 
These estimations have a certain degree of ambiguity. The approach to estimate $\beta=170$ has been used in Ref.~\cite {epl} and gives results in good agreement with the archeological data. But, differently of our case, i.e., the Earth, for the maximum carrying capacity of Easter Island, it was relatively easy to get an accurate estimation. The second range of  $\beta$ estimations based on the actual data of growing population, $\beta\simeq 700$, even if in principle could be accurate, it suffers from a limited window of time and could be affected by population fluctuations. For these reasons, we consider a range $170\leq \beta \leq 700$ in the numerical simulations.

The system is complemented by the equation describing the technological development

\begin{equation}
\frac{d}{dt}T=\alpha T\xi(t).
\label{dic}
\end{equation}
where~$T$ is the technological level that we identify with the energy consumption,~$\alpha$ a constant parameter describing the technological growth  and we may, rather optimistically, choose the value $\alpha = 0.345 $ ys$^{-1}$ following the Moore Law~\cite{moore}. Finally~$\xi(t)$ is a random variable with values~$0,1$. The time durations of the states~$0,1$ are characterized by a waiting time distribution density~$\psi(t)$. The random variable takes in account the fact that the technological development can have stops due to the economic investments on the spatial technology. 

The above model assumes that the parameters $r,\,\, \beta,\,\,r',\,\,R_c,\,\,a_0,\,\,\alpha$ have constant values over time. In this short communication we focus on the time dependence of the parameter $r$ that is directly related to the reproductive capability of the species under consideration, i.e., the human kind. In Ref.~\cite{unp}, data show that the population growth rate has been slowly decreasing at a rate $\gamma$ of $0.017 \% $per year,  and is projected to doing so in the next decades.
%\textcolor{red}{POSSIAMO DIRE CHE LA DIMINUZIONE DELLO SPERMA SI RIPERCUOTE PIU O MENO NELLA STESSA MANIERA SUL COEFFICIENTE r?} 

For such  value of  $\gamma$ we estimate   the time evolution  of $r(t)$ as

\begin{equation}
\frac{dr}{dt}=-\gamma r(t),
\label{rchan}
\end{equation}
giving an exponential behavior for $r(t)$. Note that if, in dimensionless unit, $\gamma\ll 1$, we end up into the the linear case $r(t)\approx r_0+ \gamma t$. In other words, the exponential case, for $\gamma\ll 1$, includes the constant rate case, i.e., the linear case.  Adding Eq. (5)  to Eqs. (1) and (2) we can now evolve the interaction human-forest with the addition of a time-dependent population growth-rate $r(t)$.
Running in parallel the (stochastic) evolution of the technological growth given by Eq. (4) we can estimate if the  technologic growth $T(t)$ hits the Dyson-sphere limit before
the human-forest system reaches the no-return point, after which a catastrophic  collapse in population is inevitable (see \cite{noi} for details).
Averaging over the stochastic realisations we can estimate the probability of this event occurring. 

\begin{figure}[h!]
\centering
\includegraphics[width=.55\textwidth]{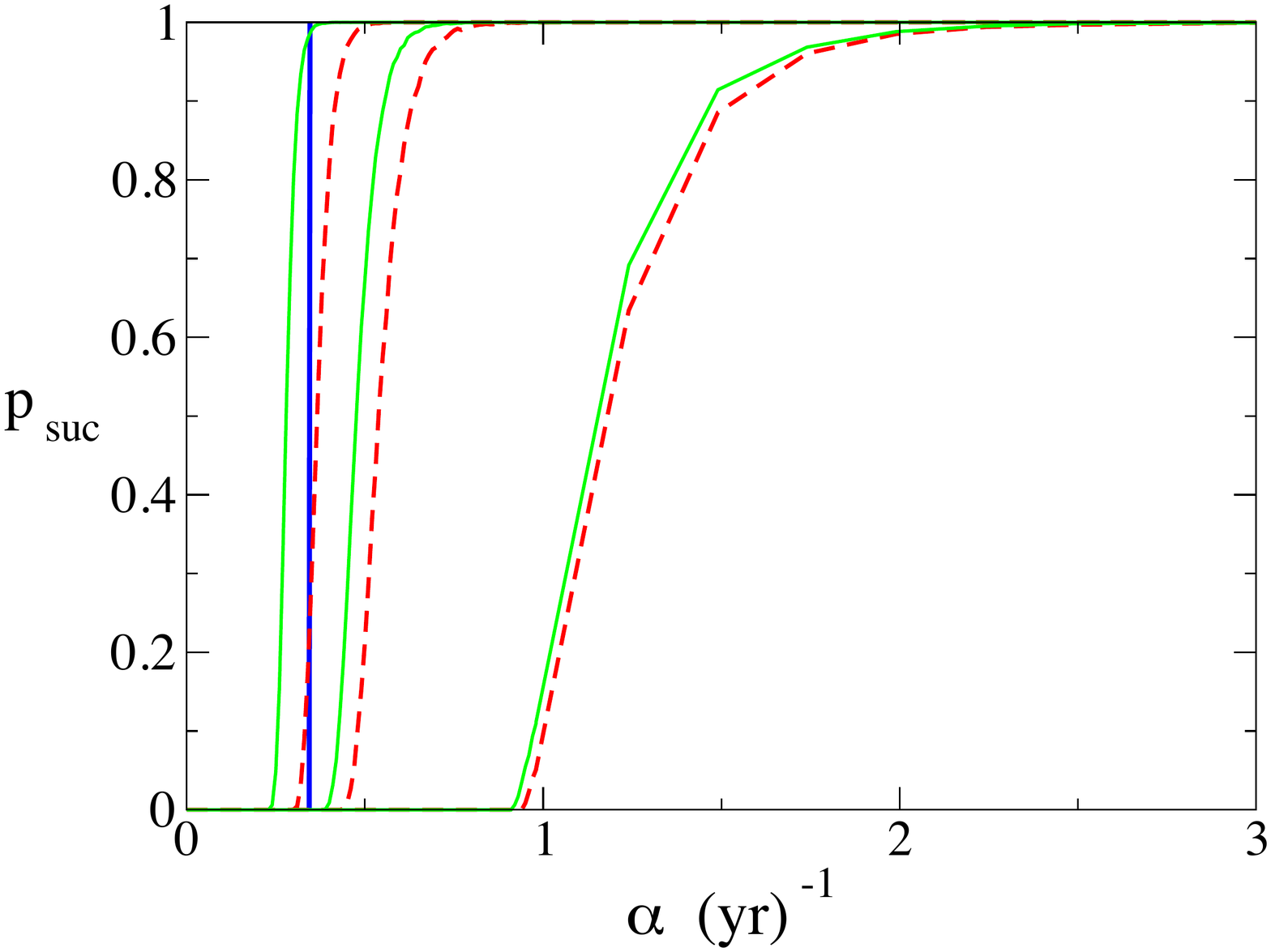}
\caption{Probability of surviving without a catastrophic collapse in population. Red dashed lines are obtained with constant $r=0.01$ years$^{-1}$. Green lines are obtained with a  time-dependent decreasing value of $r(t)$. From right to left the values of $\beta$ for the three couples of curves are $\beta=170,300,700$. The blue line indicates the Moore's law value $\alpha=0.345  $ ys$^{-1}$.
\label{fig1}}
\end{figure}
We consider several scenarios  depending on the value of the parameter $\beta$  and the time dependence of $r$. With $\beta$ ranging $170<\beta<700$ and assuming an exponential decrease  for  the rate $r(t)$. %Since the changing rate $\gamma$ is sufficiently small, we are taking into account also the linear dependence on time of $r(t)$ since $\exp[x]\approx 1+x$ for $x\ll 1$.
We obtain that for  more realistic values of $\beta <600 $ the probability of survival remains very low even if is slightly increased compared to the case of constant growth-rate $r$, confirming results obtained in \cite{noi}.
For $\beta=700$ though, we observe a dramatic increase from $10-20\%$ (where the dashed red curve intersects the blue line in Fig. 1) to more than $90\%$ probability of avoiding collapse as compared to the case with constant $r$.
Therefore, we find that only in the  remote scenario that  we were able to increase the carrying  capacity to values  $\beta> 600$  and  the technological growth-rate to the (rather optimistic)  values $\alpha \gtrapprox 0.3 $ ys$ ^{-1}$, then a decreasing population growth-rate  rate,  as stemming from recent data, would  give a  more positive outlook on our chances to avoid a catastrophic collapse in population. 
\section*{Disclosure statement}
No potential conflict of interest was reported by the authors.

\section*{ORCID}
Gerardo Aquino https://orcid.org/0000-0003-3228-7520\newline
Mauro Bologna https://orcid.org/0000-0002-6306-6897

\end{document}